\documentclass[useAMS,usenatbib]{mn2e}
\usepackage{epsfig,amsmath,amssymb}
\usepackage{graphicx}
\newcommand{\be}{\begin{equation}}    
\newcommand{\ee}{\end{equation}}
\newcommand{\beq}{\begin{eqnarray}}
\newcommand{\eeq}{\end{eqnarray}}
\newcommand{\beqn}{\begin{eqnarray*}}
\newcommand{\eeqn}{\end{eqnarray*}}

\title[Stochastic background of gravitational waves emitted by magnetars]
{Stochastic background of gravitational waves emitted by magnetars}

\author[Stefania Marassi, Riccardo Ciolfi, Raffaella Schneider, Luigi Stella \& Valeria Ferrari]{Stefania Marassi$^{1}$\thanks{E-mail:
stefania.marassi@roma1.infn.it}, Riccardo Ciolfi$^{1}$\thanks{E-mail: riccardo.ciolfi@roma1.infn.it}, 
Raffaella Schneider$^{2}$, Luigi Stella$^{3}$, Valeria Ferrari$^{1}$\\
$^{1}$Dipartimento di Fisica `G. Marconi', Sapienza Universit\`a di Roma and Sezione INFN ROMA1, Piazzale Aldo Moro 5, 00185 Roma, Italy\\ 
$^{2}$INAF/Osservatorio Astrofisico di Arcetri, Largo Enrico Fermi 5, 50125 Firenze, Italy \\
$^{3}$INAF/Osservatorio Astronomico di Roma, Via di Frascati 33, 00040 Monteporzio Catone, Italy}

\begin{document}

\date{7 September 2010}

\pagerange{\pageref{firstpage}--\pageref{lastpage}} \pubyear{2010}

\maketitle

\label{firstpage}

\begin{abstract}

Two classes of high energy sources in our galaxy are believed
to host magnetars, neutron stars whose emission results from 
the dissipation of their magnetic field.
The extremely high magnetic field of magnetars distorts their 
shape, and causes the emission of a conspicuous gravitational 
waves signal if rotation is fast and takes place around a different 
axis than the symmetry axis of the magnetic distortion. 
Based on a numerical model of the cosmic star formation history, 
we derive the cosmological background of 
gravitational waves produced by magnetars, when they are very 
young and fast spinning. We adopt different models 
for the configuration and strength of the internal 
magnetic field (which determines the distortion) as well as 
different values of the external dipole field strength (which 
governs the spin evolution of magnetars over a wide range of 
parameters). We find that the expected gravitational wave 
background differs considerably from one model to another.
The strongest signals are generated for magnetars with very 
intense toroidal internal fields ($\sim 10^{16}$~G range)
and external dipole fields of $\sim 10^{14}$, as envisaged 
in models aimed at explaining the properties of the 
Dec 2004 giant flare from SGR~1806-20. Such signals should 
be easily detectable with third generation ground 
based interferometers such as the Einstein Telescope. 

\end{abstract}


\section{Introduction}

It is well known that a variety of astrophysical processes are able to generate 
a stochastic gravitational wave background (GWB), with distinct spectral properties 
and features \citep{FMSa,FMSb,SFCFM,SFMP,RM,MSF}.
The detection of these astrophysical GWBs can provide insights into the cosmic
star formation history and constrain some of the physical properties of
compact objects, white dwarfs, neutron stars and black holes.  
Moreover, these signals may act as foreground noise for the detection
of cosmological GWBs over much of the accessible frequency spectrum. 

In this paper we consider the GWB produced by {\it magnetars}, i.e.
neutron stars with extremely high magnetic fields. Two classes 
of sources of high energy radiation in our Galaxy, 
the soft gamma repeaters (SGRs) and anomalous X-ray pulsars (AXPs), 
are believed to host magnetars, that power their emission 
through the release of magnetic field energy. 
These two classes of high-energy sources share a number of
features among which are the spin period in a fairly narrow
and long range (in the 2 - 12 s
range), the spin-down timescales ($10^4 - 10^5$~yr), the relatively 
faint persistent emission (typically $10^{34}-10^{35}$~erg/s), 
the emission of sporadic short bursts ($\ll 1$~s) with peak 
luminosities in the $10^{36}-10^{41}$~erg/s range 
(see e.g. \citealt{mer08} and references therein).

To successfully account for the observed features of both SGRs and
AXPs, the magnetar model envisages that the neutron star possesses
an internal magnetic field with strength  $B > 10^{15}$~G which
comprises both a toroidal and a poloidal component. The 
external B-field, on the contrary, is expected to be poloidal; its
dipole strength is usually inferred to be in the $10^{14}-10^{15}$~G range,
based on the observed spin-down rate (as well as other indications). 
Strong internal magnetic fields ($\sim 10^{15}$ to $10^{16}$ G) will 
induce significant quadrupolar deformations in the neutron star
structure; these may generate a detectable gravitational 
wave signal, if their symmetry axis is not aligned with the spin 
axis (see e.g. \citet{C} and references therein).

Using population synthesis methods to evaluate the initial period
and the magnetic field distributions of magnetars, \citet{RFP2} computed 
the GWB due to a magnetar population, assuming a  purely-poloidal
magnetic field configuration both inside of the star and 
in the magnetosphere.
They found that the largest signal is obtained for
a type I superconductor neutron star model, and that the resulting 
closure energy density peaks at $\Omega_{\rm GW} \sim 10^{-9}$ around
1.2~kHz;  this is well below the sensitivity of the first
generation of detectors, but it might be an interesting target for 
future detectors, e.g. the Einstein
Telescope\footnote{http://www.et-gw.eu/}, as discussed in detail in
\citet{RM}.

In the present work, we reconsider the GWB generated by magnetars,
by using the cosmic star formation rate density
evolution predicted by the numerical simulation of \citet{TFS}, and
adopting several magnetar models recently proposed in the
literature.  As a first example, we use the purely poloidal
configurations discussed in \citet{RFP2}. We then consider {\it
twisted torus} configurations, recently discussed in
\citet{CFGP,CFG} and in \citet{ff};  in these models  the poloidal component of the
magnetic field extends throughout the  star and in the magnetosphere,
whereas the toroidal component is confined to a torus-shaped
region inside the star. Finally, we consider the model in  \citet{SOIV}
(see also \citealt{dal09}), namely
an internal field configuration dominated by the toroidal component 
with strength $\sim2\cdot10^{16}$~G (core-averaged value), and a poloidal 
field of ordinary strength ($10^{14}-10^{15}$ G).
Our main purpose here is to assess how the uncertainties
related to the internal magnetic field strength and its
configuration affect the resulting GW signal and whether
next-generation detectors will have the potential to reveal 
such a signal and shed light on the properties 
of magnetars. 

The plan of the paper is as follows. In Section \ref{sectionsfr} we
briefly describe the numerical simulation performed by \citet{TFS}
that we use to predict the cosmic star formation rate evolution and
the corresponding magnetar birthrate. In Section
\ref{sectionsinglesource} we sketch out the main features of the
single source spectrum which describe the gravitational emission of
a single magnetar, and introduce the different magnetar models
considered in this work. In Section \ref{sectionGWB}, we present
the resulting density parameter of the GWB, $\Omega_{\rm GW}$, and
discuss its detectability. In Section \ref{secwobble} we 
discuss the effect of the wobble angle on the generated background
and its relevance for the detection of the signal. 
Finally, in Section \ref{sectionconclusion} we draw our conclusions.

Throughout our work we have adopt a $\Lambda$CDM cosmological
model with parameters $\Omega_M=0.26$, $\Omega_\Lambda=0.74$,
$h=0.73$, $\Omega_b=0.041$, in agreement with the three-year WMAP
results \citep{S}. 

\section{The magnetars birth rate evolution}
\label{sectionsfr}

Following \citet{MSF}, we use the cosmic star formation rate density
evolution predicted by the numerical simulation of \citet{TFS}.  For
the present study, we consider the formation rate of Population
II stars only; for these we adopt a Salpeter Initial Mass
Function (IMF) $\Phi(M)\propto M^{-(1+x)}$ with $x=1.35$ (normalized
between $0.1 - 100 M_{\odot}$), in regions of the Universe which
have been already polluted by the first metals and dust grains
(\citealt{SFNO,SFSOB}; \citealt{OTSF}). 

\citet{RFP} and \citet{PP} derived the statistical properties of
highly magnetized neutron stars (B$\geq 10^{14}$ G) by 
using population synthesis methods; they showed that neutron stars
born as magnetars represent 8-10\% of the total simulated population
of neutron stars.
Here we assume a fraction f$_{\rm MNS}=$ 10\%; we further assume
that magnetar progenitors have masses in the $8 M_{\sun}-40
M_{\sun}$ range.  It should be noted that the mass range of 
magnetar progenitors is still debated (see for
instance \citealt{FW, DFK}).  However, the range we consider is
sufficiently large to include the proposed evolutionary models.

\begin{figure} 
\includegraphics[width=8.5cm,angle=360]{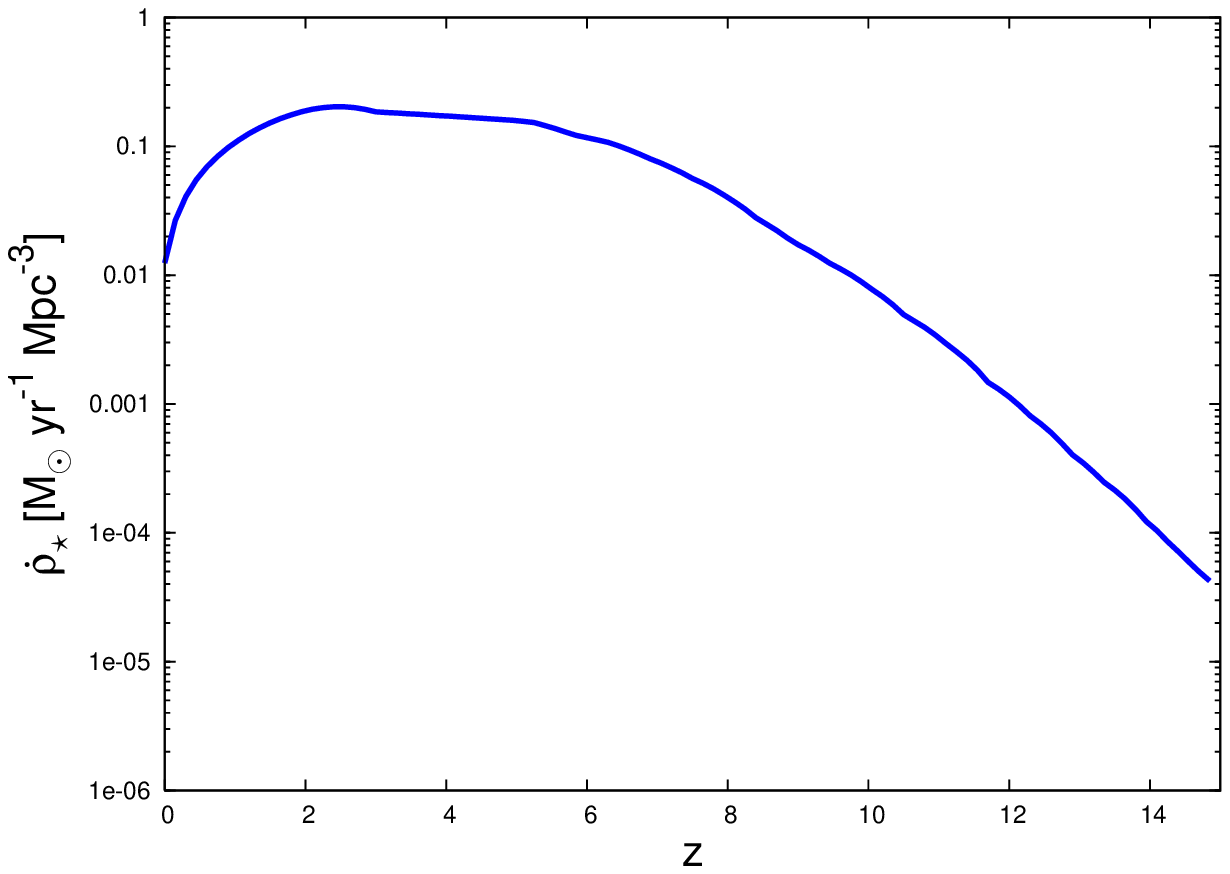}
\includegraphics[width=6.0cm,angle=270]{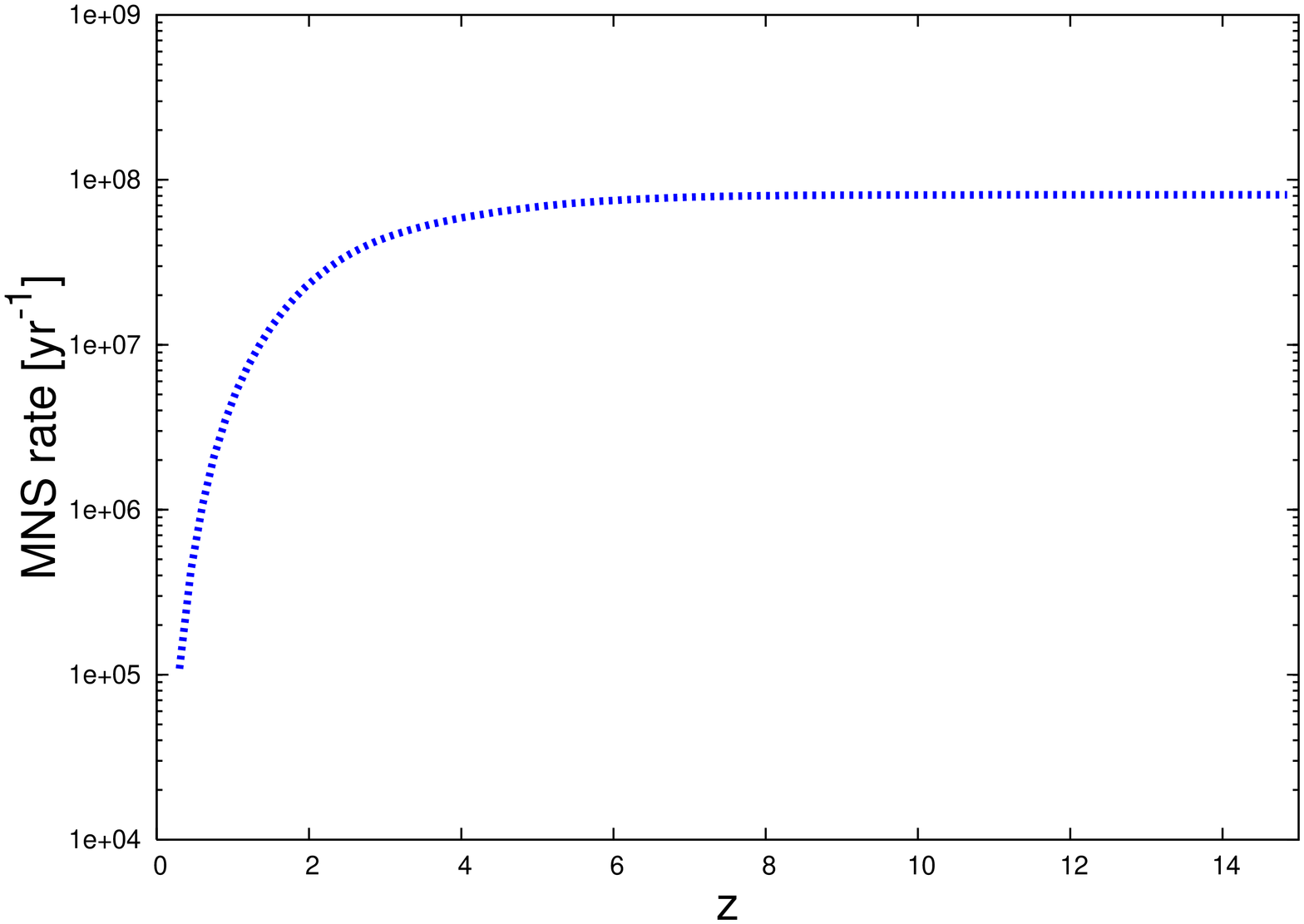}
\caption{{\it Top panel:} redshift evolution of the comoving star
formation rate density.  {\it Bottom panel:} redshift evolution 
of the number of magnetars formed per unit time.} \label{sim:sfr} \end{figure}

The top panel of Fig.~\ref{sim:sfr} shows the redshift evolution of
the cosmic star formation rate density inferred from the
simulation\footnote{The results shown in Fig.~\ref{sim:sfr} refer to
the fiducial run in \citet{TFS} with a box of comoving size $L=10
h^{-1}$~Mpc and $N_{\rm p} = 2 \times 256^3$ (dark+baryonic)
particles.}. 

The number of magnetars formed per unit time out to a given redshift
$z$ can be computed by integrating the cosmic star formation rate
density, $\dot{\rho_\star}(z)$, over the comoving volume element, while
restricting the integral over the stellar IMF in the proper range of
progenitor masses; that is 
\be R_{\rm MNS}(z) =f_{\rm MNS} \int_0^z
dz' \frac{dV}{dz'}\frac{\dot{\rho}_{\star}(z')}{(1+z')}\int_{8
M_\odot}^{40 M_\odot} dM \Phi(M), 
\ee
where the factor $(1+z)$ at the denominator accounts for the time-dilation
effect, and the comoving volume element can be expressed as 
\beq &&dV=4 \pi r^2  \left(\frac{c}{H_0}\right) \epsilon(z) dz \\
&&\epsilon(z) = \left[\Omega_M(1+z)^3+\Omega_\Lambda
\right]^{-\frac{1}{2}}.\nonumber  \label{comvol} \eeq 
The result is shown in the bottom panel of Fig.~\ref{sim:sfr}. 

\section{Magnetars as gravitational wave sources}
\label{sectionsinglesource}

The gravitational wave energy spectrum emitted by a single source is
\be
\frac{dE_{GW}}{df_e}=
\dot{E}_{GW}\bigg{|}\frac{df_e}{dt}\bigg{|}^{-1} ,
\label{eq:gwspectrum}
\ee
where $f_e$ is the emission frequency. The gravitational wave
luminosity of a neutron star with spin axis forming a wobble angle
$\alpha$ with the magnetic axis is composed of two contributions,
one at the spin frequency $\nu_R$, one at its double $2\nu_R$; it
can be written as
\be 
\dot{E}_{GW}= \frac{2G}{5c^5}I^2\epsilon_B^2
\omega^6 \sin^2{\alpha} (\cos^2\alpha+16\sin^2{\alpha})  \, ,
\label{lum} 
\ee 
where $\epsilon_B$ is the star quadrupolar ellipticity induced
by the magnetic field, $\omega$ is
the angular velocity and $I$ is the moment of inertia.  The term
$\sin^2{\alpha} \cos^2\alpha$ is from the emission  component 
at the spin frequency, the term $16\sin^4{\alpha}$ from the component 
at twice the spin frequency.

In strongly magnetized neutron stars the quadrupolar deformation is
determined essentially by the  magnetic field configuration and
strength\footnote{Fast
rotation will also induce a non-negligible deformation, 
which, being symmetric with respect to the spin axis, does not 
contribute to the gravitational wave emission.}.  Moreover, it depends on the
equation of state of matter in the stellar interior.

The star loses rotational energy mainly due to electromagnetic
radiation and gravitational wave emission 
(we shall neglect other effects, e.g. relativistic winds;
\citealt{dal09}). According to the well known vacuum 
dipole radiation model, the energy loss rate due to dipole radiation is given by 
\be 
|\dot{E}_{ROT}^{dip}|= \frac{1}{6}\frac{B_p^2
R^6}{c^3}\omega^4\sin^2{\alpha}, \label{diprad} \ee 
where $B_p$ is the field strength at the magnetic poles, and $R$ 
the stellar radius.  The total spin-down rate  obtained from
Eqs.~(\ref{lum}) and (\ref{diprad}) is 
\beq |\dot{\omega}|&=&|\dot{\omega}^{dip}|+|\dot{\omega}^{GW}|
 \\ &=&\frac{1}{6}\frac{B_p^2
R^6}{Ic^3}\omega^3\sin^2{\alpha}+\frac{2G}{5c^5}I\epsilon_B^2
\omega^5 \sin^2{\alpha} (1+15\sin^2{\alpha}) .\nonumber
\label{omegadot}
\eeq
Using the above quantity and remembering that the first
term of the gravitational wave luminosity given in Eq.
(\ref{lum}) is emitted at  $f_e=\nu_R=\omega/2\pi$, 
while the second at $f_e=2\nu_R$, we can compute 
the single source emission spectrum according to Eq. (\ref{eq:gwspectrum}).   

The choice of the initial spin period $P_0$ for the neutron star  population
sets an upper limit on the frequency ranges where the two components of the 
GW emission contribute to the gravitational wave background:
the emission at $\nu_R$ contributes to frequencies below $1/P_0$, that 
at $2\nu_R$ to frequencies below $2/P_0$. Therefore,
for  $f_e < 1/P_0$ the gravitational wave background
has both contributions, while  for $1/P_0 < f_e < 2/P_0$ 
the only contribution comes from the 
emission at $2\nu_R$.
Hence, the terms to be considered 
when computing the spectral energy density of the background are:\\
for $f_e < \frac{1}{P_0}$ ,
\beq 
&&\frac{dE_{GW}}{df_e}= \frac{32\pi^4 G}{5c^5}I^2\epsilon_B^2
f_e^3 \nonumber \\ && \times \Bigg{\{} \cos^2{\alpha} \left[
\frac{B_p^2 R^6}{6Ic^3}+\frac{8\pi^2 G}{5c^5}I \epsilon_B^2 f_e^2
(1+15\sin^2{\alpha}) \right]^{-1} \nonumber \\ &&+ \sin^2{\alpha}
\left[ \frac{B_p^2 R^6}{6Ic^3}+\frac{2\pi^2 G}{5c^5}I \epsilon_B^2
f_e^2 (1+15\sin^2{\alpha}) \right]^{-1} \Bigg{\}} ;
\label{alphagen} 
\eeq 
for $\frac{1}{P_0} < f_e < \frac{2}{P_0}$ ,
\beq
&&\frac{dE_{GW}}{df_e}= \frac{32\pi^4 G}{5c^5}I^2\epsilon_B^2
f_e^3 \nonumber \\ && \times 
 \sin^2{\alpha}
\left[ \frac{B_p^2 R^6}{6Ic^3}+\frac{2\pi^2 G}{5c^5}I \epsilon_B^2
f_e^2 (1+15\sin^2{\alpha}) \right]^{-1} ;
\label{alphagen_bis}
\eeq
for $f_e > \frac{2}{P_0}$ , 
\beq
&&\frac{dE_{GW}}{df_e} = 0\; .\nonumber
\eeq

If we assume $B_p=10^{14}-10^{15}$ G, $R\sim 10$ km, $I\sim10^{45}$
g cm$^2$ and $f_e\lesssim 1$ kHz, we see that even for quadrupole
ellipticities as large as $10^{-4}$ the term $\frac{B_p^2
R^6}{6Ic^3}$ is much larger than $\frac{8\pi^2 G}{5c^5}I\epsilon_B^2
f_e^2$; in this case the contribution of gravitational wave
emission to the spin-down is negligible. As shown in the following, 
this holds in most of the cases we consider. It is worth noting that, 
when $\alpha\neq 0$ and $\frac{B_p^2R^6}{6Ic^3}\gg\frac{8\pi^2 G}{5c^5}I\epsilon_B^2f_e^2$,
for $f_e < 1/P_0$ the dominant term in Eq. (\ref{alphagen}) is 
\be \frac{dE_{GW}}{df_e}= \frac{32\pi^4 G}{5c^5}I^2\epsilon_B^2
f_e^3 \left( \frac{B_p^2 R^6}{6Ic^3}\right)^{-1} \, ,
\label{alphagen2} \ee 
which does not depend on the wobble angle $\alpha$\footnote{It should be noted 
that if we remove the term $\sin^2\alpha$ in Eq.~(\ref{diprad}) (see for instance \citealt{OG}),
the $\alpha$ dependence in Eq.~(\ref{alphagen2}) is preserved.}.
However, for $\frac{1}{P_0} < f_e <\frac{2}{P_0}$, the dominant term is
\be \frac{dE_{GW}}{df_e}= \frac{32\pi^4 G}{5c^5}I^2\epsilon_B^2
f_e^3 \sin^2\alpha  \left( \frac{B_p^2 R^6}{6Ic^3}\right)^{-1} \, ,
\label{alphagen2_bis} 
\ee
and it depends on $\alpha$. 

It should be stressed that in general the wobble angle depends on
time.  The misalignment of magnetic and rotation axes causes, in the
neutron star frame, the free precession of the angular velocity
around the magnetic axis with period $P_{prec}\simeq
P/|\epsilon_{B}|$, where $P$ is the spin period \citep{J,JA}.  The
star internal viscosity  damps such precessional motion and reduces
the wobble angle towards the aligned configuration ($\alpha=0$), if
the star has an oblate shape ($\epsilon_B>0$), whereas it increases
$\alpha$ towards the orthogonal configuration $\alpha=\pi/2$
(``spin-flip''), if the shape is prolate ($\epsilon_B<0$)\citep{JPB,C}.  
The second case is more favourable for gravitational wave emission.

The timescale of the process is given by
$\tau_\alpha=nP_0/\epsilon_B$, where $P_0/\epsilon_B$ is the 
initial precession period and $n$ is the expected number 
of precession cycles in which the process takes place;  
estimates for slowly rotating neutron stars indicate that 
$n\sim 10^2-10^4$ \citep{AS}, however the value of $n$ is actually
unknown \citep{C}.  The evolution of the misalignment angle is
relevant for our analysis only if the associated timescale,
$\tau_\alpha$, is short compared to the spin-down timescale,
$\tau_{sd}$; conversely, if $\tau_\alpha \gg \tau_{sd}$ the process
takes place when the source is no longer an efficient gravitational
wave emitter.

In principle, an accurate estimate of the GWB should account for (i) a 
proper distribution of the initial wobble angles for the magnetar population,
and (ii) the evolution of such misalignment with time. This kind of analysis 
would, however, be affected by the wide uncertainties on 
both $\tau_{\alpha}$ and the initial angle 
distribution. Nevertheless, as shown in Section \ref{secwobble}, we found that, 
from the point of view of the GWB detectability, the value of $\alpha$ and 
its eventual change with time do not significantly affect the results. 
Since our main interest is focused on detection prospects, we proceed here 
with the simplifying assumption that all magnetars are born with $\alpha=\pi/2$ 
and that the misalignment evolution is ineffective. Then, in Section \ref{secwobble}, 
we will consider the effects of a generic wobble angle. 

An essential input for $dE_{GW}/df_e$ is the magnetic field strength
at the pole, $B_p$: it determines the electromagnetic spin-down rate
and, depending on the model, it may also affect the
stellar deformations.  To be representative of the entire
population, the value of $B_p$ should be chosen as a suitable
average. Such an average is uncertain at present; however 
the values of $B_p$ inferred from AXPs and SGRs  
lie in the $10^{14}-10^{15}$ G range (\citealt{mer08}).  
Our choice here is to span this range by studying its 
two extremes, $B_p=10^{14}$ and $10^{15}$ G.  As we shall
see, this translates into an uncertainty in the results that is 
negligible
in comparison with the uncertainties associated to our poor
knowledge of the internal field configuration.

As we have seen, the overall GW emission depends on $P_0$, 
the initial spin period. Following \citet{RFP2} we set $P_0=0.8$ ms.  
For $\alpha=\pi/2$ this gives $f_e^{max}=2/P_0=2500$ Hz. 
If $\alpha<\pi/2$, part of the gravitational wave 
emission is at the spin frequency and the corresponding 
contribution has a frequency cutoff $f_e^{max}=1/P_0=1250$ Hz. 
The chosen value of $P_0$ implies a very fast spinning newborn 
neutron star, but still consistent 
with the believed range of neutron star spin rates at birth. We 
remark that, according to current scenarios of magnetar formation, 
strongly magnetized neutron stars are those that are born with 
periods of the order of ms, much faster than ordinary 
pulsars (\citealt{dt92}). At the end of Section \ref{detectsub} 
we will sketch the effect of assuming lower initial spin frequencies.

In the following, we compute the gravitational wave background
assuming different magnetic field models.
\subsection{Purely poloidal magnetic field} \label{secpoloidal}
The first two field configurations we consider describe a strongly
magnetized neutron star endowed with a purely poloidal magnetic
field, and have been used to evaluate the corresponding GWB in \citet{RFP2}.  

It is well known that poloidal fields tend to make the star oblate
($\epsilon_B>0$), while toroidal fields deform the star in a
prolate shape ($\epsilon_B<0$). Therefore, in the case we consider
in this Section, ellipticity is always positive. Here we compute the
gravitational wave emission according to Equations (\ref{alphagen}) and (\ref{alphagen_bis}),
setting, as in \citet{RFP2},  $\alpha=\pi/2$ and assuming that
the viscous evolution of the wobble angle
(which tends towards $\alpha=0$ in the case of positive
ellipticities, thus reducing GW emission) is slower than the 
spin-down of the star. For a given 
poloidal configuration of the internal B-field, 
the GW output is maximized under the above assumption.

The numerical inputs to compute $dE_{GW}/df_e$ are $B_p$,
$\epsilon_B$, $I$ and $R$.  Following \citet{KOK}, we write
the quadrupolar ellipticity as
\be \epsilon_B =g\frac{B_p^2 R^4}{GM^2}, \ee
where $M$ is the mass of the star and the value of the dimensionless
(deformation) parameter $g$ accounts for the magnetic field geometry
and the EOS.  As in \citet{RFP2}, we consider two models with
$g=13$ (Model A) and $g=520$ (Model B), respectively. The first
model refers to an incompressible fluid star and a dipolar magnetic
field \citep{F}; similar values are obtained in relativistic models
based on polytropic equations of state \citep{KOK}.  Model B
describes a scenario in which the neutron star core is a
superconductor of type I, implying that the internal magnetic field
is confined to the crustal layers \citep{BG}.  This scenario gives
much stronger deformations (see also \citealt{COL}). 

As previously discussed, we adopt two different values of $B_p$:
$10^{14}$ and $10^{15}$ G.  The other parameters are fixed as
follows: $R=10$ km, $M=1.4$ M$_{\odot}$ and $I=10^{45}$ g cm$^2$.

\subsection{Twisted torus configurations}
In order to account for the observed features of AXPs and SGRs,
the magnetar model envisages that 
the internal magnetic field is a
mixture of poloidal and toroidal components
\citep{dt92,td93,td95}. 
A strong internal toroidal components is expected
to form as a result of differential rotation 
shortly after the birth of neutron star. The 
magnetic energy stored in this component
provides the energy reservoir to power the bulk
of the magnetar emission throughout its lifetime,
including short bursts and giant flares 
\citep{wt06}. 
The poloidal component of the B-field (or 
at least part of it) extends outside the star,
giving rise to a magnetosphere. 
However, the detailed configuration of the magnetic 
field is presently unknown.
Recent studies of the evolution of
strongly magnetized stars in Newtonian gravity
indicate that a particular magnetic field
configuration, the so-called {\it twisted torus}, 
is a quite generic outcome of dynamical simulations and,
due to magnetic helicity conservation, appears to be stable on
dynamical time-scales \citep{BS1,BS2,BN}.
In this model,  the poloidal field extends throughout
the star and the exterior, whereas the toroidal field is confined
in a torus-shaped region inside the star.

Here we consider  twisted torus
equilibrium configurations of strongly magnetized neutron stars in
the framework of general relativity \citep{CFGP,CFG}.  The magnetic
field includes contributions from higher
multipoles ($l>1$) coupled to the dipolar ($l=1$) field which is usually
assumed; an argument of minimal energy is adopted in order to establish the
relative weights of the different multipoles as well as the
relative strength of toroidal and poloidal fields.  In this model
the poloidal fields dominate over the toroidal ones.  As a
consequence, these stars have always an oblate shape
\citep{CFGP,CFG,LJ}.  In order to
account for the dependence on the equation of state, two different EOSs are
employed, named APR2 and GNH3, which span a realistic range of
compactness \citep{CFG}.

As in the case of purely poloidal fields, we compute the
gravitational wave emission spectrum according to 
Equations (\ref{alphagen}), (\ref{alphagen_bis}) and assuming 
$\alpha = \pi/2$.  For a given magnetic field strength
($B_p=10^{14}$ and $10^{15}$ G) and stellar mass ($M=1.4$
M$_{\odot}$), the model provides the ellipticity, radius and 
moment of inertia of the neutron star; we have
\be \epsilon_B \simeq k \left( \frac{B_p}{10^{15} \text{ G}} \right)^2
\cdot10^{-6}, 
\ee \noindent 
with $k=9$ $(4)$, $R=14.19$ $(11.58)$ km,  and
$I=1.82$ $(1.33)\cdot10^{45}$ g\,cm$^2$, for the EOS GNH3 (APR2).

\subsection{Toroidal-dominated magnetic field model}
The last model we consider is based on the hypotesis of very 
strong toroidal fields inside the star. We assume a magnetic field 
configuration with an internal toroidal field of $\sim2\cdot 10^{16}$ G 
(core-averaged value), in addition to a poloidal field of ordinary strength 
($10^{14}-10^{15}$ G).
\citet{SOIV} showed that toroidal field strengths of this
order are needed to explain the time-integrated emission of magnetars 
as inferred from the extremely bright giant flare that took place 
on 2004 December 27 from SGR~1806-20 (which liberated an energy of 
$5 \times 10^{46}$~erg).  Giant flares of this magnitude could 
result from large-scale rearrangements of the core magnetic 
field or instabilities in the magnetosphere \citep{td01,lyu03}.  
Such a huge toroidal field would induce prolate deformations 
as strong as $\epsilon_B\simeq-6.4\cdot 10^{-4}$ \citep{C}. Consequently, a newly-born
fast spinning magnetar is expected to emit a strong gravitational signal
whose frequency, initially in the $0.5 -2$~kHz range, decreases over 
a timescale of days (and whose strain correspondingly decreases too).
This signal should be observable with the Advanced Virgo/LIGO class detectors up 
the distance of the Virgo cluster \citep{dal09}.  
The deformation associated to poloidal fields, whose strength is fixed by 
the choice of $B_p$,  
tends to oppose the above deformation; however in the cases 
considered here ($B_p=10^{14}$ and $10^{15}$ G) the corresponding 
correction is negligible (of the order of $10^{-4}-10^{-2}$, respectively).
The physical inputs we use in Equations (\ref{alphagen}) and (\ref{alphagen_bis}), 
in addition to $\alpha=\pi/2$, $\epsilon_B$ and $B_p$, are $R=10$~km and $I=10^{45}$g\,cm$^2$. 

The gravitational wave emission predicted by the present model with
$B_p=10^{14}$ G could be regarded as an upper limit among the
different magnetar models (excluding exotic scenarios), as in this
case there occurs the most favorable combination of magnetic
fields: an extremely strong toroidal field dominates the deformation
while the lower value of the poloidal field strength results in a
slower spin-down. 
In addition, the shape of the star is prolate and the evolution of the wobble 
angle $\alpha$ leads the axis of the magnetically-induced
deformation towards the orthogonal configuration, 
resulting in stronger gravitational wave emission. 
 
\begin{figure*} \centering \begin{minipage}{176mm} \begin{center}
\includegraphics[width=6cm,angle=270]{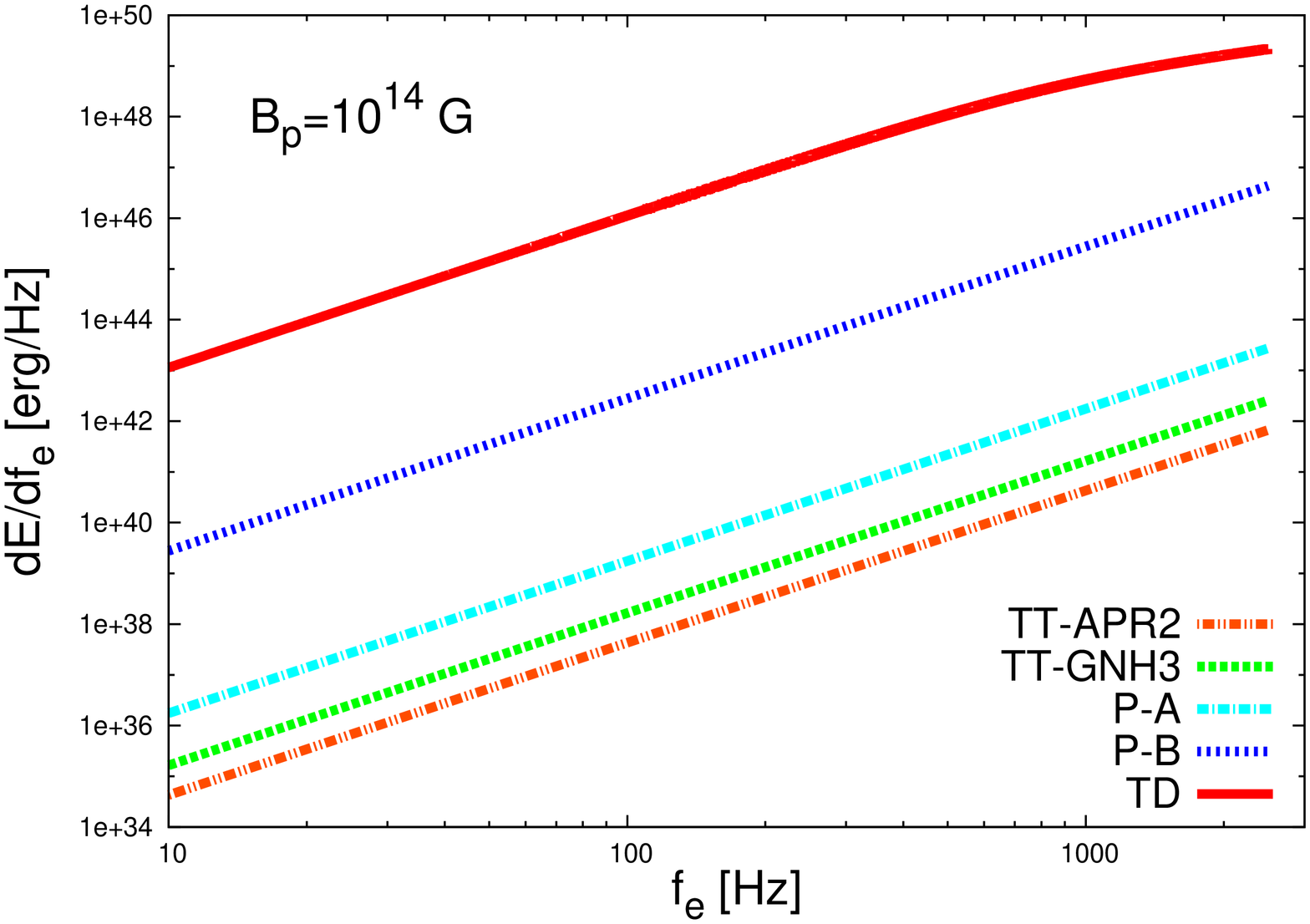}
\includegraphics[width=6cm,angle=270]{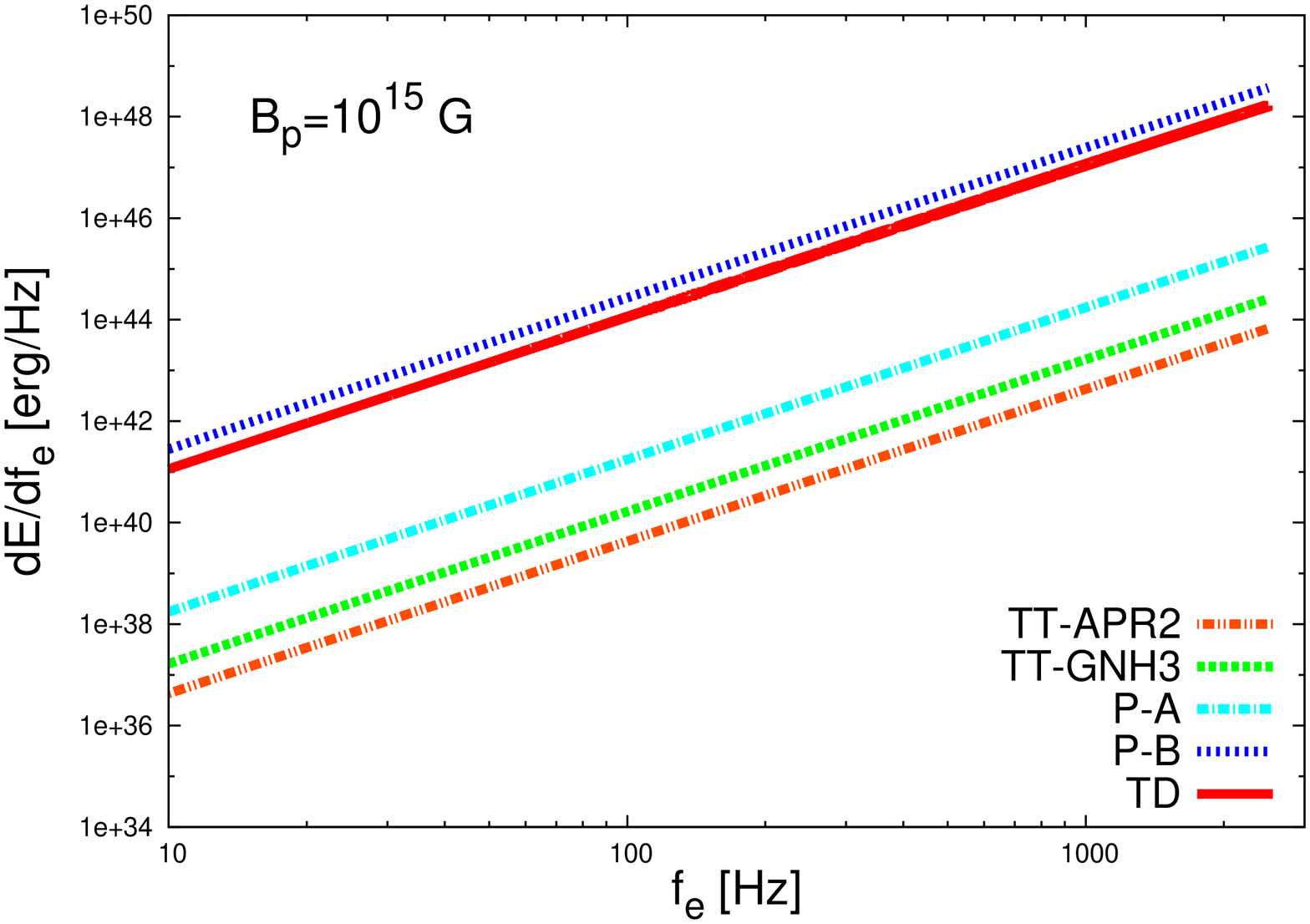}
\end{center} \caption{Spectral gravitational energy emitted by a
single source as a function of the emitted frequency, for the
different models we consider: P-A and P-B stand for the purely
poloidal models A and B, TT are the twisted torus model predictions
for the two EOS considered (APR2 and GNH3), TD indicates the
toroidal-dominated model. {\it Left panel}: $B_p=10^{14}$ G; {\it
right panel}: $B_p=10^{15}$ G.} \label{singlesourceall}
\end{minipage} \end{figure*}

\subsection{Gravitational wave emission spectrum}
\label{modelpredictions}

We present here the gravitational wave spectrum
emitted by the models illustrated above.  In Fig.
\ref{singlesourceall} we plot $dE_{GW}/df_e$ as a function of the
emitted frequency $f_e$ for the two purely poloidal models A and B
(hereafter P-A and P-B), the twisted torus model (hereafter TT) with
the two equations of state considered (APR2, GNH3), and the
toroidal-dominated model (hereafter TD).  In the left (right) panel
we assume $B_p=10^{14}$ G ($B_p=10^{15}$ G); as already discussed,
these values define a likely range for $B_p$.

The first important indication which emerges from Fig. \ref{singlesourceall} 
is that the uncertainty related to the different magnetar models is always
much higher (3-5 orders of magnitude) than the spread associated
to the adopted range of $B_p$.  
Let us now focus on the $B_p=10^{14}$ G case (left panel).  The TD
model is by far the most favorable for gravitational wave emission,
having the optimal combination of strong deformation and slow
electromagnetic spin-down.  The second strongest emission is obtained
with the P-B model, where large deformations are achieved even for
this lower field strength.  The emission predicted by the P-A model
is lower by more than three orders of magnitude, due to the
difference in the $g^2$ factor appearing in the expression of
$dE_{GW}/df_e$.  The two TT models are expected to give even 
weaker signals; they differ for the assumed EOS and, as expected, 
the one which gives less (more) compact stars, GNH3 (APR2), 
is associated to stronger (weaker) deformations and gravitational wave emission.

If we consider higher external poloidal fields
($B_p=10^{15}$ G, right panel) the picture changes.  For the P-A,
P-B and TT models the value of $B_p$ controls both the gravitational
wave luminosity, which scales as $B_p^4$ ($\epsilon_B\propto
B_p^2$), and the spin-down rate, which has the electromagnetic
contribution proportional to $B_p^2$ plus a very small correction
due to gravitational wave emission.  As a result, $dE_{GW}/df_e
\propto B_p^2$, which translates in a factor 100 increase from $B_p
= 10^{14}$ to $10^{15}$ G. Conversely, in the TD model the
deformation is determined by the dominant toroidal field in the
stellar interior (with poloidal field corrections up to $\sim 1$\%
for $10^{15}$ G), and an increase in $B_p$ only results in a higher
electromagnetic spin-down and in a smaller overall gravitational wave
emission.  As long as the electromagnetic spin-down dominates over
the gravitational wave spin-down, $dE_{GW}/df_e$ is {\it reduced} by
a factor of 100 from $B_p = 10^{14}$ to $10^{15}$ G.  The final
result is that when $B_p=10^{15}$ G, the prediction of TD and P-B
models are comparable. 

\begin{figure}
\includegraphics[width=6cm,angle=270]{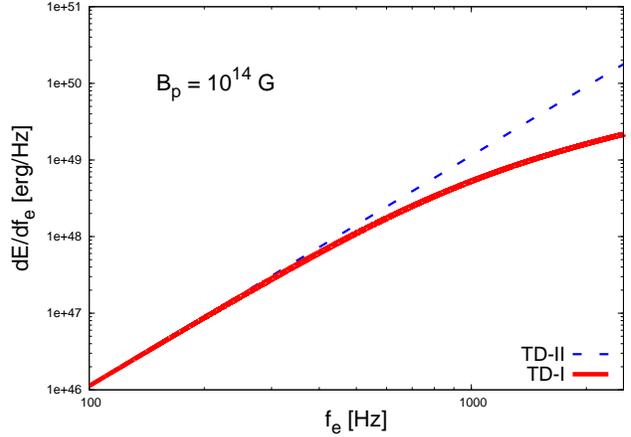}
\caption{Spectral gravitational energy emitted by a single source according 
to the toroidal-dominated model as a function of the emitted frequency.
TD-I is the same as TD in the left panel of Fig.
\ref{singlesourceall}; in TD-II (dashed line) the contribution of
gravitational wave emission to the  spin-down is
neglected.} \label{split} \end{figure}

It is worth noting that in all the considered models the contribution
given by the gravitational wave emission to the spin-down is negligible,
with the exception of the early time evolution in the 
TD model with $B_p=10^{14}$ G (hereafter
TD$_{14}$).  This is shown in Fig. \ref{singlesourceall}, where the energy
spectra are linear in logarithmic scale, reflecting the behaviour
$dE_{GW}/df_e\propto f_e^3$, while the TD$_{14}$ model is characterized by
a lower emission level at high frequency, due to a non-negligible gravitational 
wave spin-down.  This effect is even more evident in Fig.~\ref{split}, where we
compare the gravitational wave spectrum for the same TD model shown in the
left panel of Fig.~\ref{singlesourceall} with a TD model where
gravitational wave spin-down is neglected (dashed line). It is clear that the
gravitational wave contribution starts to be relevant at $f_e \sim 300$
Hz. For all the other models we have discussed, this contribution becomes relevant
at much higher emission frequencies.

\section{Gravitational Wave Background from magnetars}
\label{sectionGWB}
In this Section we compute the GWBs produced by the different 
magnetar models presented in the previous Section. 
Following \citet{MSF}, the spectral energy density 
of the GWB can be written as 
\be
\frac{dE}{dS df dt}=\int^{z_f}_{0} \int^{M_f}_{M_i} dR(M,z)
 \big{<}\frac{dE}{dS df}\big{>} \, ,
\label{gwbk}
\ee
where $dR(M,z)$ is the differential source formation rate
\be
dR(M,z)=\frac{\dot{\rho}_\star(z)}{(1+z)}\frac{dV}{dz}\Phi(M)dMdz,
\label{bkrate}
\ee
and $\big{<}\frac{dE}{dS df}\big{>}$ is the locally measured average 
energy flux emitted by a source at distance $r$. For 
sources at redshift $z$ it becomes  
\be 
\big{<}\frac{dE}{dS df}\big{>}=\frac{(1+z)^2}{4\pi
d_L(z)^2}\frac{dE_{GW}}{df_e}[f(1+z)] \, , 
\label{singspec} 
\ee 
where $f=f_e(1+z)^{-1}$ is the redshifted emission frequency $f_e$, and $d_L(z)$
is the luminosity distance to the source.
\begin{figure*} 
\centering 
\begin{minipage}{176mm} \begin{center}
\includegraphics[width=6cm,angle=270]{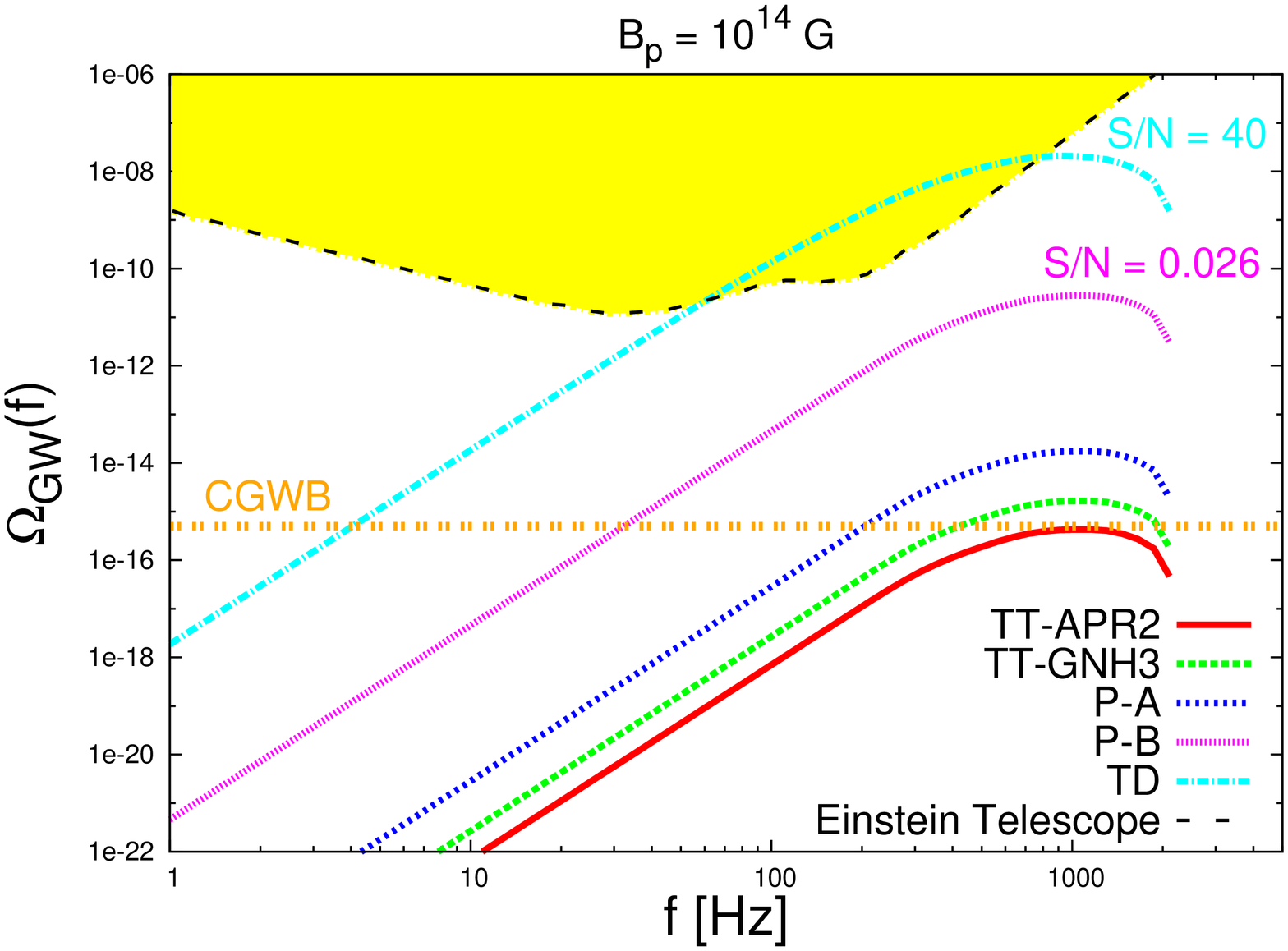}
\includegraphics[width=6cm,angle=270]{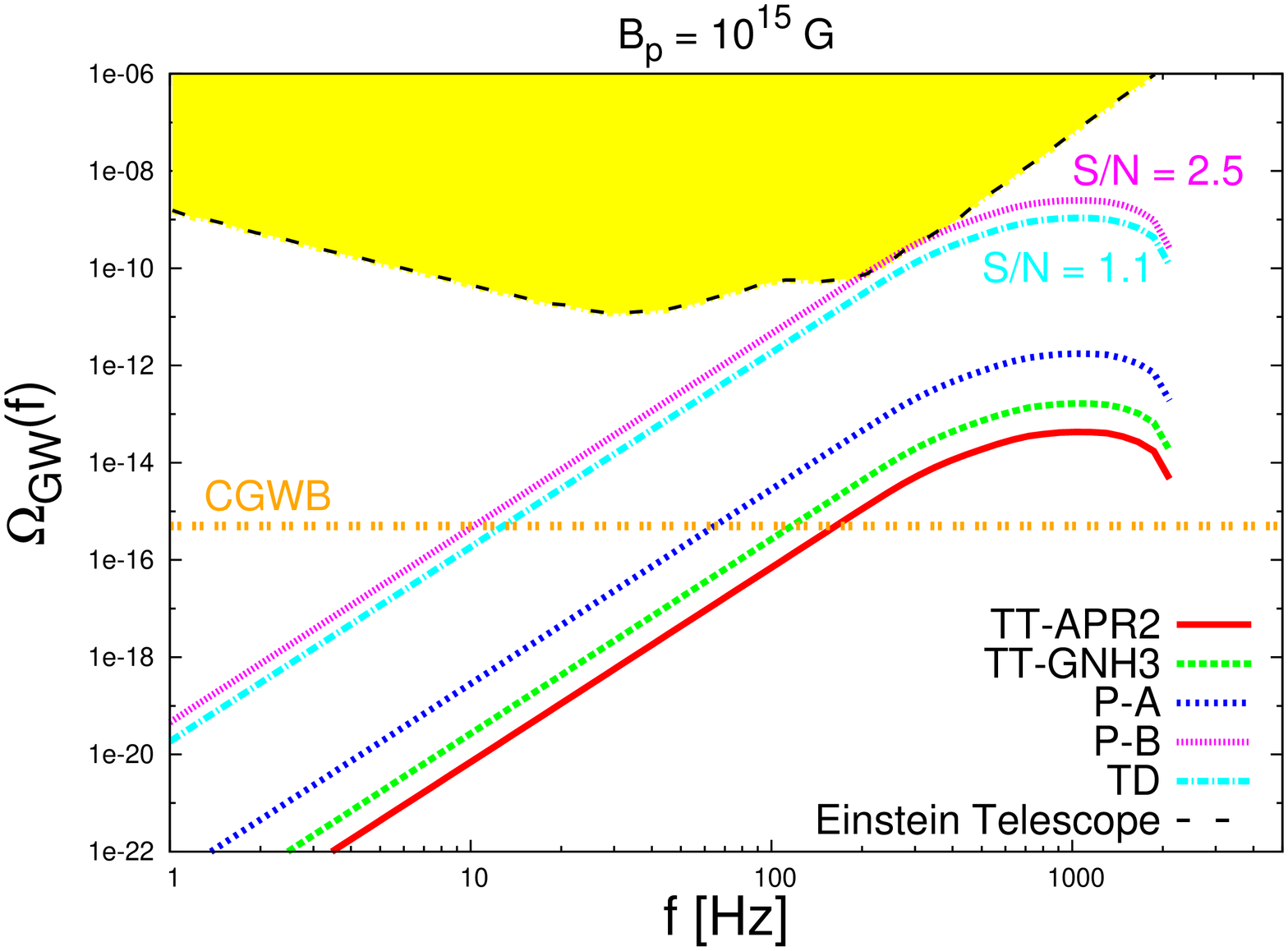}
\end{center} \caption{The predicted closure energy density ($\Omega_{\rm
GW}$) as a function of the observational frequency, for the different
magnetar models discussed in this paper: 
P-A and P-B stand for purely poloidal models A
and B, TT are the twisted torus model predictions for the two EOS
considered (APR2 and GNH3), TD is the toroidal-dominated model.  {\it Left
panel}: $B_p=10^{14}$ G; {\it right panel}: $B_p=10^{15}$ G.  In both
panels the shaded region indicates the foreseen sensitivity of the
Einstein Telescope, and the horizontal dotted line (CGWB) is the upper limit on
primordial backgrounds generated during the Inflationary epoch. A given background is 
detectable by the Einstein Telescope if the corresponding signal-to-noise 
ratio is larger than the detection threshold $S/N=2.56$ (see text).}
\label{detect} \end{minipage} \end{figure*}

It is customary to describe the GWB by a
dimensionless quantity, the closure energy density
 $\Omega_{\rm GW}(f) \equiv {\rho_{cr}}^{-1}(d\rho_{gw}/dlog f)$, 
which is related to the spectral energy density by the equation
\be
\Omega_{\rm GW}(f)=\frac{f}{c^3\rho_{cr}}\left[\frac{dE}{dS df dt}\right] ,
\ee
where $\rho_{cr}=3H_0^2/8\pi G$ is the cosmic critical density.

In Fig.~\ref{detect}, we show $\Omega_{\rm GW}$ as a function of the
observational frequency for the different magnetar models. Differences in
the predicted stochastic backgrounds reflect differences in the
corresponding single magnetar emission spectrum.  The maximum amplitude is
always achieved around 1 kHz: in the left panel, it ranges from $\sim
4\cdot 10^{-16}$ to $\sim 2\cdot 10^{-8}$, while in the right panel the
range is $\sim 4\cdot 10^{-14} - 2\cdot 10^{-9}$.  The higher value is
obtained with the TD model in the first case, and with the P-B model in
the second case; the lower value is given in both cases by the TT-APR2
model.  In both panels, model predictions are compared with the foreseen
sensitivity of the Einstein Telescope (shaded region) and with the upper
limit to primordial backgrounds generated during the Inflationary epoch
(horizontal dotted line labelled CGWB). The latter contribution 
is estimated from Eq.~(6) of \citet{T} assuming a tensor/scalar ratio 
of $r=0.3$ and no running spectral index of tensor perturbations \citep{KKMR1,KKMR2}.

It is clear from the figure that the background generated by magnetars
in the kHz region is higher than the primordial background, independently of
the specific magnetar model considered; in addition, specific magnetar
models lead to a cumulative signal which is potentially detectable by the
Einstein Telescope. A more quantitative assessment of the detectability is
reported in the following Section.   

\subsection{Detectability}
\label{detectsub}

The gravitational wave signal produced by the magnetar population
can be treated as continuous. Indeed, if $\Delta \tau_{\rm gw}$ is
the average time duration of a signal produced by a single magnetar,
and $dR(z)$ is the number of sources formed per unit time at redshift
$z$, the duty cycle $D$ out to redshift $z$, defined as
\be
D(z)= \int_0^z dR(z) \Delta\tau_{\rm gw}(1+z) \, ,
\label{duty}
\ee
satisfies the condition\footnote{If we take $\Delta \tau_{\rm
gw}=\tau_{sd}$ we have always $D$ higher than $10^{3}$.} $D \gg 1$.
Consequently, the stochastic signal appears
in the detector outputs as a time-series noise which, by the central limit
theorem, is expected to have a Gaussian-normal distribution function.
In this case, as suggested by \citet{AR,RM}, the optimal detection
strategy is to cross-correlate the output of two (or more) detectors,
assumed to have independent spectral noises.

The optimized $S/N$ for an integration time $T$ is given by \citet{AR},
\be
\left(\frac{S}{N}\right)^2=\frac{9H^4_0}{50\pi^4}T\int^{\infty}_0 df \frac{\gamma^2(f)\Omega^2_{\rm GW}(f)}{f^6P_1(f)P_2(f)} \, ,
\label{snr}
\ee
\noindent
where $P_{1}(f)$ and $P_{2}(f)$ are the power spectral noise 
densities of the two detectors, and $\gamma$ is the normalized 
overlap reduction function, characterizing the loss of sensitivity 
due to the separation and the relative orientation of the detectors. 

The sensitivity of detector pairs is given in terms of the 
minimum detectable amplitude for a flat spectrum $\Omega_{\rm MIN}$ 
($\Omega_{\rm MIN}= const$) defined as
\beq
\Omega_{\rm MIN}&=&\frac{1}{\sqrt T}\frac{10\pi^2}{3H^2_0}\left[\int^\infty_0 df \frac{\gamma^2(f)}{f^6P_1(f)P_2(f)}\right]^{-1/2} \nonumber\\
&&\cdot(\rm{erfc^{-1}}(2\alpha)-\rm{erfc^{-1}}(2\gamma)) \, , \\\nonumber
\eeq  
where $T$ is the observation time, $\alpha$ the false alarm rate, $\gamma$ the detection rate
and $\rm{erfc^{-1}}$ the complementary error function (for more details see \citealt{AR}).

If we consider the cross-correlation of two detectors with the sensitivity
of the Einstein Telescope (T. Regimbau, private communication), we get
$\Omega_{\rm MIN}=1.13\cdot 10^{-11}$ for an integration time $T$ of one
year, a false alarm rate $\alpha=10\%$ and a detection rate $\gamma=90\%$;
these values, inserted in Eq.~(\ref{snr}), lead to a detection threshold
$S/N$ of 2.56.  

A given background is detectable by the Einstein Telescope if the 
corresponding signal-to-noise ratio given by Eq.~(\ref{snr}) is larger 
than the detection threshold. For instance, the predicted $\Omega_{\rm GW}$ 
for the purely poloidal model P-B with $B_p=10^{15}$~G gives $S/N=2.49$, that 
is slightly smaller than such threshold; consequently, there is 
no chance to detect this signal. Conversely, in the most optimistic
magnetar model $TD_{14}$ (toroidal-dominated model with $B_p=10^{14}$~G)
we obtain $S/N=40$, a very promising value.  This result leads to the
conclusion that third-generation gravitational wave detectors, such as the
Einstein Telescope, hold the potential to reveal the cumulative GW signal
from magnetars in the universe. 

It is worth noting that the above results refer to the assumed 
initial spin period of $P_0=0.8$ ms. A higher value would lead to a 
lower frequency cutoff and, consequently, to a weaker GWB. For the 
$TD_{14}$ model, for example, the detection threshold $S/N=2.56$
corresponds to $P_0=5.2$ ms. Hence the GWB would still be detectable 
up to this value. 

\section{Wobble angle effects}
\label{secwobble}

So far we have assumed a constant misalignment $\alpha=\pi/2$
between the spin and the magnetic axis, in which case the GW signal 
is emitted only at twice the spin frequency $f_e=2\nu_R=\omega/\pi$.
For a generic misalignment, we have also the emission at the spin frequency. 

\begin{figure}
\includegraphics[width=6cm,angle=270]{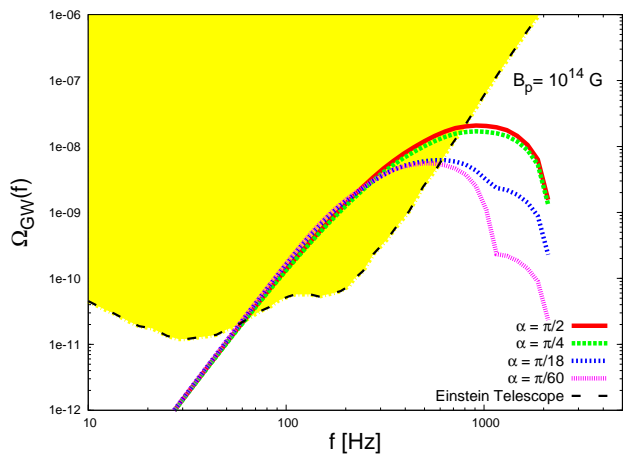}
\caption{The TD$_{14}$ background is plotted for different wobble angles, 
spanning the range $\pi/60-\pi/2$, and compared with the Einstein Telescope sensitivity.}
\label{figwobble}
\end{figure}

We now focus on the model TD$_{14}$ and explore the consequences of $\alpha<\pi/2$. 
In this model the stellar deformation induced by the magnetic field are larger;
being the most optimistic model for gravitational wave emission, this case allows 
to clearly  show the effects of the wobble angle on detectability.

In Fig. \ref{figwobble} we compare the sensitivity of the Einstein
Telescope with the background generated by model TD$_{14}$ for different
(constant) values of the wobble angle.  The plot clearly shows that when
$\alpha \gtrsim \pi/4$ there is a single dominant contribution with the
frequency cutoff at 2500 Hz, while for smaller angles there is a dominant
contribution with $f^{max}=1250$ Hz and a secondary contribution with
lower amplitude extending up to $f^{max}=2500$ Hz. Similar effects hold
for the alternative magnetar models which have been presented in the
previous Sections. A difference between the models potentially 	arises if the
timescale for the evolution of the wobble angle is short compared to the
spin-down timescale (see Section \ref{sectionsinglesource}): in this case, the
star rapidly tends (i) to the orthogonal configuration for the TD model, 
thus increasing gravitational wave emission, and (ii) to the aligned
configuration,  for models P-A, P-B and TT models, thus decreasing the emission.

As shown in Fig. \ref{figwobble}, for the TD$_{14}$ model 
(as well as for the other models) 
the gravitational wave backgrounds corresponding to different 
wobble angles exhibit significant differences at large frequencies, 
approximately above $\sim 800$~Hz, where the Einstein Telescope 
sensitivity is too low even for this model;
therefore, the signal detectability  is only marginally affected.
Variations in the signal-to-noise ratio are at most 2-3\% in the TD$_{14}$
case, and if the gravitational wave background is weaker (e.g. for
$B_p>10^{14}$ G) the effects on the S/N are even smaller.

We can conclude that the initial value of $\alpha$ and its evolution 
in time do not have significant effects on the GWB detectability with 
the Einstein Telescope.

\section{Conclusions}
\label{sectionconclusion}

In this paper, we estimated the GWB produced by magnetars.  We used a cosmic
star formation history obtained from a numerical simulation performed by
\citet{TFS} and assume that 10\% of stellar progenitors with masses in the
range $8 - 40 M_{\odot}$ lead to magnetars with magnetic field  strength in
the $10^{14}-10^{15}$ G range.

Since our present understanding of the physical properties of magnetars is
still poor (the internal structure of the magnetic field and initial spin 
frequency being among the major uncertainties), we have
explored the consequences for gravitational wave emission of different
magnetar models proposed in the literature. 

Our analyis shows that different models produce a spread in the resulting
gravitational wave emission which is much higher than that produced by
adopting different values for the magnetic field strength. 
In particular, we find that:
\begin{itemize}
\item Toroidal-dominated models, with an internal toroidal field of $\sim
2 \cdot 10^{16}$~G and an external poloidal field of
$10^{14}$~G, proposed by Stella et al. (2005) to explain the 2004 giant
flare from SGR 1806-20, generate the largest gravitational wave
background, which could be detected in the frequency range between
$\sim 50$ and $\sim 600$~Hz by third generation gravitational wave detectors
such as the Einstein Telescope.  Using correlated analysis of Einstein
Telescope outputs, the estimated  signal-to-noise ratios
could be as high as 40.
\item When larger poloidal fields, $10^{15}$~G, are considered, the largest
gravitational background is generated by magnetar models  with purely
poloidal fields, and a superconductor type-I core; 
in this case, the internal magnetic field is confined to the
crustal layers, leading to strong deformations. Since deformations are
produced by the internal toroidal field, Toroidal-dominated models are
less effective because the increase in the poloidal field strength leads
to a higher electromagnetic spin-down and to a lower gravitational wave
emission.
\item A comparison between the estimated magnetar GWB and the upper limit 
to the primordial background predicted by Inflationary scenarios (the 
horizontal dotted line in Fig. \ref{detect} labelled CGWB) shows that, 
for the models of magnetar we consider, the magnetar GWB is always 
larger than the primordial background  in some region of frequency 
(the only exception is the twisted-torus model TT-APR2 with $B_p=10^{15}$~G). 
For instance, for the toroidal dominated models TD$_{15}$ and TD$_{14}$ 
this is true, respectively, for $f \gtrsim 13$ Hz and $f \gtrsim 4$ Hz.
Thus, the GWB generated by magnetars may act as a limiting 
foreground for the future detection of the primordial background 
even at frequencies as low as few tens of Hz.
\end{itemize}

We have also investigated the consequences on the resulting gravitational
wave background of different values for the misalignement angle between
the rotation and magnetic field axes. We find the largest effects to be at
high frequencies, above $\sim800$~Hz; thus the detectability of the
largest backgrounds with the Einstein Telescope is only marginally
affected, with fractional variations of the signal-to-noise ratios of at
most 2-3\%. 


\section*{Acknowledgments}

We thank Tania Regimbau for useful suggestions and discussions 
and for providing us the ET correlated sensitivity curve 
and $\gamma$ function.

Stefania Marassi thanks the Italian Space Agency (ASI) for the support.
This work is funded with the ASI CONTRACT I/016/07/0.

This work was also supported by CompStar, a Research Networking 
Programme of the European Science Foundation.


\label{lastpage}

\end{document}